\documentclass{emulateapj}
\usepackage{epsfig}
\usepackage{multirow}

\begin{document}

\title{HST/WFC3 Observations of Low-Mass Globular Clusters AM 4 and Palomar 13: Physical Properties and Implications for Mass Loss}

\author{Katherine M. Hamren\altaffilmark{1,2}, Graeme H. Smith\altaffilmark{1}, Puragra Guhathakurta\altaffilmark{1}, Andrew E. Dolphin\altaffilmark{3}, Daniel R. Weisz\altaffilmark{4},  Abhijith Rajan\altaffilmark{5}, Carl J. Grillmair\altaffilmark{6}}

\altaffiltext{1}{Department of Astronomy and Astrophysics, University of California Santa Cruz, 1156 High Street, Santa Cruz, CA  95064, USA}
\altaffiltext{2}{{\tt khamren@ucolick.org}}
\altaffiltext{3}{Raytheon, 1151 E. Hermans Road, Tucson, AZ 85756, USA}
\altaffiltext{4}{Department of Astronomy, Box 351580, University of Washington, Seattle, WA 98195, USA}
\altaffiltext{5}{School of Earth and Space Exploration, Arizona State University, 781 E Terrace Road, Tempe, AZ 85287, USA}
\altaffiltext{6}{Spitzer Science Center, California Institute of Technology, Mail Stop 220-6, Pasadena, CA 91125, USA}

\date{}

\begin{abstract}

We investigate the loss of low-mass stars in two of the faintest globular clusters known, AM 4 and Palomar 13 (Pal 13), using HST/WFC3 F606W and F814W photometry. To determine the physical properties of each cluster --- age, mass, metallicity, extinction, present day mass function (MF) --- we use the maximum likelihood color-magnitude diagram (CMD) fitting program MATCH and the Dartmouth, Padova and BaSTI stellar evolution models. For AM 4, the Dartmouth models provide the best match to the CMD and yield an age of $> 13$ Gyr, metallicity log $Z/Z_\odot = -1.68 \pm 0.08$, a distance modulus $(m-M)_V = 17.47 \pm 0.03$ and reddening $A_V = 0.19 \pm 0.02$. For Pal 13 the Dartmouth models give an age of $13.4 \pm 0.5$ Gyr, log $Z/Z_\odot = -1.55 \pm 0.06$, $(m-M)_V = 17.17 \pm 0.02$ and $A_V = 0.43 \pm 0.01$. We find that the systematic uncertainties due to choice in assumed stellar model greatly exceed the random uncertainties, highlighting the importance of using multiple stellar models when analyzing stellar populations. Assuming a single-sloped power law MF, we find that AM 4 and Pal 13 have spectral indices $\alpha = +0.68 \pm 0.34$ and $\alpha = -1.67 \pm 0.25$ (where a Salpeter MF has $\alpha = +1.35$), respectively. Comparing our derived slopes with literature measurements of cluster integrated magnitude ($M_V$) and MF slope indicates that AM 4 is an outlier. Its MF slope is substantially steeper than clusters of comparable luminosity, while Pal 13 has a MF in line with the general trend. We discuss both primordial and dynamical origins for the unusual MF slope of AM 4 and tentatively favor the dynamical scenario. However, MF slopes of more low luminosity clusters are needed to verify this hypothesis.
\end{abstract}


\maketitle

\section{Introduction}

The last decade has seen a surge of theoretical and observational evidence that mass loss plays a significant role in the evolution of globular clusters (GCs). GCs lose mass via two channels: mass loss from individual stars, and loss of the stars themselves. The latter process --- the focus of this paper --- can occur via two-body relaxation processes or as a result of external effects like tidal stripping and  tidal shocking. Two-body relaxation is  well documented by \textit{N}-body simulations \citep[e.g.][]{Fall2001,DErcole2008}, while the presence of features like tidal tails \citep[e.g.][]{Grillmair1995, Leon2000,Fellhauer2007,NO2010,Sollima2011} in GCs constitutes observational evidence for mass loss by external forces like tidal stripping.

As GCs evolve, low-mass stars are more likely to be lost than massive stars. The underlying cause of this is mass segregation, in which massive stars migrate towards the cluster center while less-massive stars migrate to the outskirts. Whether this mass segregation is primordial \citep{Baumgardt2008} or the result of energy equipartition \citep{PZ2001,Baumgardt2003}, the low-mass stars in the outer reaches of the GC are more prone to evaporation and stripping. In addition, tidal shocking by the Milky Way's bulge or disk will heat a GC, causing it to lose (the predominantly low-mass) stars from its outskirts \citep{OSC1972,CKS1986}.
 
The preferential loss of low-mass stars is reflected in a GC's main-sequence luminosity function (MSLF) and in the slope of its mass function (MF). \citet{Pryor1991} were some of the first to indicate that the MSLF of a cluster that has been losing stars will be flatter at the faint end (i.e. more bottom light) than a cluster that has not. This theory was tested by \citet{GrillmairSmith2001} on the faint GC Palomar 5, whose tidal tails are clear evidence for mass loss due to tidal stripping \citep{Odenkirchen2001, Grillmair2006}. Grillmair \& Smith found that while there was not a sharp cutoff in Pal 5's MSLF, it was considerably more bottom light than the MSLFs of $\omega$ Cen \citep{deMarchi1999} or M55 \citep{Paresce_&_DeMarchi2000}. Similar arguments have been made regarding the slope of a cluster's MF, where a flat MF is often taken as evidence of severe tidal stripping \citep[as in the case of NGC 6218,][]{DeMarchi2006}.

Considerable attention has been paid to the likely loss of stars from bright, massive GCs, in large part due to the role of mass loss in the development of multiple stellar populations \citep[see][and references therein]{Conroy2012}. Less attention has been paid to mass loss in low-mass, low-surface brightness clusters. In addition to Pal 5, several particularly faint GCs show flat MSLFs/MFs; Whiting 1 \citep{Carraro2007}, Pal 1 \citep{Rosenberg1998}, and Pal 4 \citep{Frank2012}. Contrasting these MSLFs to those of more massive GCs like M92 \citep{Paust2007}, 47 Tuc \citep{Monkman2006}, or NGC 2419 \citep{Bellazzini2012}, which show steadily increasing MSLFs, it appears that low-mass GCs are particularly susceptible to dissolution and indeed may have been much more massive in the past.

In this paper we compare MSLFs and MF slopes of two particularly low-mass GCs, Pal 13 and AM 4, derived from the first Hubble Space Telescope (HST) data obtained for either cluster. As part of our investigation of the MSLF/MF, we also present age, distance, metallicity and extinction estimates for the two clusters. This paper is organized as follows. In Section \ref{obs} we describe the HST observations and data reduction. In Section \ref{analysis} we describe our analysis of the color magnitude diagrams (addressing field stars, binary stars, and blue straggler stars), and in Section \ref{results} we present our determinations of physical parameters (age, distance, metallicity, MSLF, mass function slope) of each cluster. We discuss the implications of these results in Section \ref{disc}.

\section{Observations and Data Reduction}\label{obs}

\subsection{WFC3 Imaging}

The observations were obtained with the HST Wide Field Camera 3 (WFC3) under program number GO-11680 (PI: Smith) during Cycle 17. Our program encompassed four full orbits, one per filter per cluster. HST images are typically under-sampled; however, this was remedied by making use of the exquisite pointing stability of the telescope and dithering the observations. Each orbit was made up of four dithered exposures, resulting in total exposure times of $\sim2460$s per filter per cluster. 

\subsection{Photometry}\label{phot}

We performed PSF-fitting photometry using the WFC3 module of the DOLPHOT photometry package \citep{dolphin2000}. DOLPHOT performs photometry on non-drizzled images (i.e. bias, flat and dark calibrated \_FLTs), using the drizzled images for alignment only. The reduction procedure first masks bad pixels and cosmic rays. It then splits the four dithered \_FLT frames into their 16 component ``chips" (digital arrays corresponding to the four amps per \_FLT) and, using an analytical Tiny Tim PSF \citep{Krist1995}, performs PSF photometry simultaneously on all 16 chips. With this method, a ``single" measurement is a combination of measurements at the same sky location on all 16 chips rather than the average of 16 independently photometered images. The F606W drizzled images were used for alignment. After applying aperture corrections, DOLPHOT used published zeropoints\footnote{Available online at http://www.stsci.edu/hst/wfc3/phot\_zp\_lbn} to convert instrumental magnitudes to the VEGAMAG system. The output catalogs were cleaned of non-astrophysical and poorly measured sources by using the DOLPHOT sharpness and crowding metrics: $|$sharp$| < 0.1$ and $crowding < 0.25$. ``Good" detections passed both tests, while ``bad" detections failed one or both. Detections were also required to have SNR$\ge 5$ in both filters.

\begin{figure}[]
\begin{center}
\includegraphics[width=3.5in]{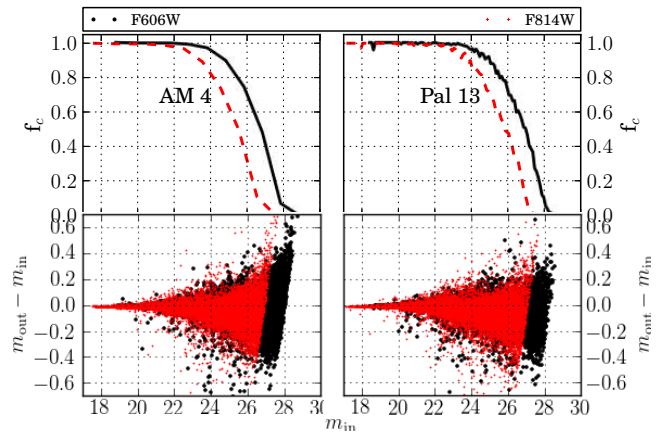}
\caption{The upper panels show the completeness fractions for both filters, and the lower panels show the residuals between input and output magnitudes (F606W in black, F814W in red). The differences between filters and clusters are minimal. For all panels, the x axis is input magnitude in the VEGAMAG system. }
\label{fig:completeness} 
\end{center}
\end{figure}

We modeled the incompleteness and photometric uncertainties in our data by utilizing the artificial star test capabilities of DOLPHOT. We generated $\sim56000$ artificial stars in each cluster, added to the image and photometered one at a time to eliminate the possibility of artificial inducing crowding and blending. The resulting completeness fractions ($f_{\rm c}$) are shown in the top panels of Figure~\ref{fig:completeness}. The 50\% completeness limit is reached at $m_{\rm F606W} = 27.2$ mag, $m_{\rm F814W} = 26.0$ mag, at which point our magnitude uncertainties are $\sim 0.1$ mag. The residuals between the input and recovered magnitudes of the artificial stars are shown in the bottom panels. A quick look at the residuals confirms our assumption that crowding and blending are not a serious problem in these clusters, as they are distributed symmetrically around $m_{\rm out} - m_{\rm in} = 0$. If a substantial fraction of stars were blended together we would expect the recovered stars to be brighter than they were when inputted and the distribution of the residuals would be shifted towards brighter magnitudes. The completeness fractions are similar for both clusters and both filters.

\begin{figure*}[ht]
\begin{center}
\includegraphics[width=6.5in]{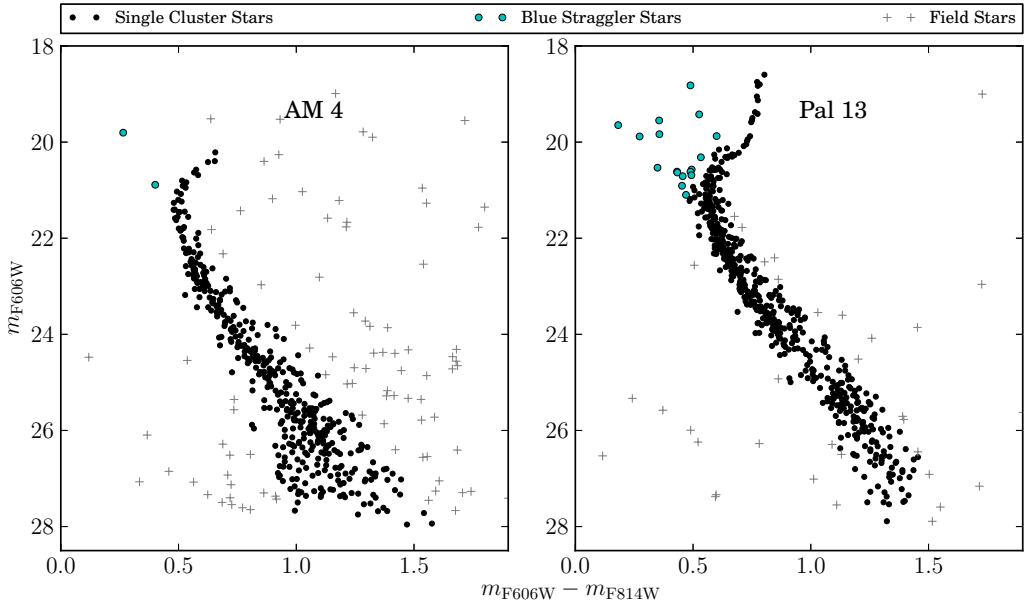}
\caption{The final CMDs of AM 4 (left panel) and Pal 13 (right panel). The stars are color-coded by type: grey crosses represent likely field stars (see Section \ref{Field}) blue circles show likely blue straggler stars (Section \ref{BSS}), and black circles represent cluster members.}
\label{fig:CMD} 
\end{center}
\end{figure*}

\section{CMD Analysis}\label{analysis}

The CMDs of AM 4 and Pal 13 are shown in Figure~\ref{fig:CMD}. The different populations of stars --- field stars, cluster stars and blue stragglers --- are highlighted, and will be discussed in the subsequent sections. The CMD of AM 4 includes 435 cluster stars, and the CMD of Pal 13 includes 640 cluster stars.

\subsection{Field Star Contamination}\label{Field}

\begin{figure}[b]
\begin{center}
\includegraphics[width=3in]{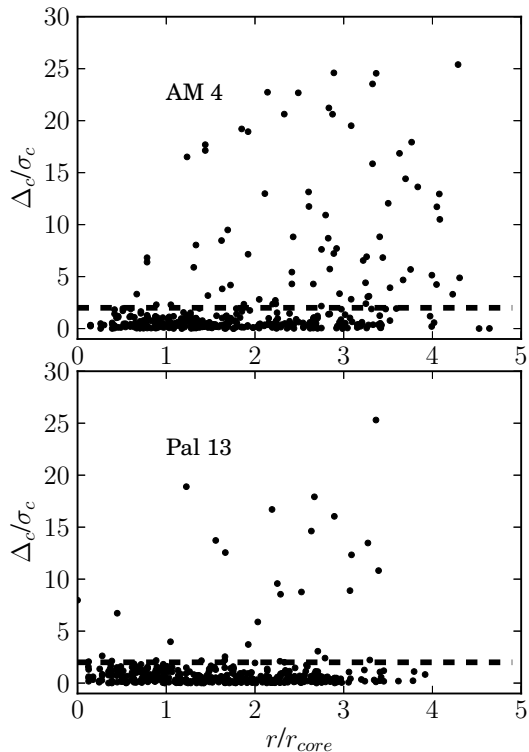}
\caption{The top panel shows the normalized distance from the cluster center vs normalized distance in color from the empirical ridge-line for every star in AM 4. The dashed line illustrates the $2\sigma_{c}$ limit, which we use to define the main-sequence envelope. The bottom panel shows the same for Pal 13.}
\label{fig:deltaSpace} 
\end{center}
\end{figure}

Field star contamination is fairly minimal at the galactic latitudes of these two clusters (Pal 13: $b = -42.70$, AM 4: $b = +33.51$). However, since the clusters have only $\sim 500$ stars, even a small number of foreground contaminants could alter the slope of the MSLF. Lacking proper motion or parallel field data, we deal with these contaminants in two ways: we define a main sequence envelope to exclude all stars whose position on the CMD rules out cluster membership, and we use the Besan\c{c}on galaxy model \citep{Robin2003} to statistically address the stars that fall within that envelope. 

To define a main-sequence envelope, we used plots of each cluster's radial and color distributions. We defined the radial distributions ($r/r_{\rm core}$) to be the distance from each star to the cluster center normalized by the cluster core radius. The cluster centers and $r_{\rm core}$ were taken from the 2010 edition of the Harris Globular Cluster Catalog \citep{Harris2010}. The color distribution ($\Delta_{\rm c}$/$\sigma_{\rm c}$) was defined as the distance in color space from the empirical MS ridge line normalized by the width of the main-sequence at that magnitude. The width of the main sequence was determined by rectifying the CMD (subtracting the color of the empirical ridge line from the color of each star), splitting the rectified CMD into five bins, and then fitting a Gaussian to the distribution of color in each bin. We defined the width of the main sequence to be the standard deviation of that Gaussian. To determine width as a function of magnitude, we fit a curve to the standard deviations in each bin. The width of the main sequence as a function of $m_{\rm F606W}$ in AM 4 was well fit by an exponential function, while in Pal 13 it was fit by a quadratic.

\begin{figure*}[t]
\begin{center}
\includegraphics[width = 6in]{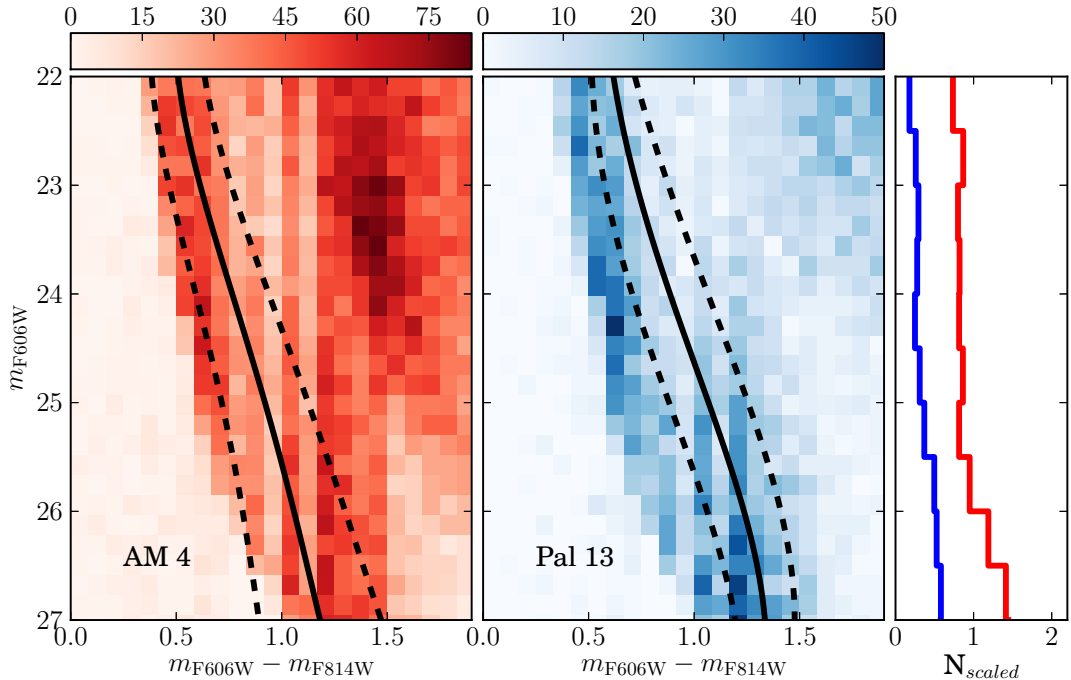}
\caption{The left panel shows the Hess diagram of the Besan\c{c}on model of the Milky Way in a 3600 arcmin$^2$ region (an area deliberately chosen to be much greater than the WFC3 field of view) centered on AM 4. The Hess diagram is over-plotted with our empirical cluster fiducial (solid line) and the $2\sigma_{c}$ envelope defining the main sequence (dashed lines). The center panel shows the same for Pal 13. The right panel shows the number of Besan\c{c}on stars per magnitude bin that fall within the main sequence envelopes, scaled by the WFC3 field of view of 7.3 arcmin$^2$. For both clusters, the contamination is uniform across the main sequence and less than two stars in any given magnitude bin. }
\label{fig:Hess}
\end{center}
\end{figure*}

Figure~\ref{fig:deltaSpace} shows $\Delta_{\rm c}$/$\sigma_{\rm c}$ versus $r$/$r_{\rm core}$. The main sequence is clearly visible as the dense population of stars along the line of $\Delta_{\rm c}$/$\sigma_{\rm c} < 1$, while the field stars occupy a range of radii and colors. Using this plot, we define our envelope to be $\Delta_{\rm c}$/$\sigma_{\rm c} = 2$ (shown in Figure~\ref{fig:deltaSpace} by the dashed lines). For brevity, we will refer to this limit as $2\sigma_c$.

To estimate the fraction of field stars per magnitude bin within the $2\sigma_c$ envelope, we used the Besan\c{c}on model of the Milky Way. We used the Besa\c{c}on model to generate artificial 3600 arcmin$^2$ fields centered on each cluster. Beginning with a field much larger than the WFC3 field of view allows us to mitigate the effects of small number statistics. 

To compare these model stars to our data, it was necessary to convert them from the Johnson-Cousins magnitude system to the VEGAMAG system. We used the findings of \citet{Bellazzini2012}, who compared the WFC3 photometry of more than 150 stars to their Johnson-Cousins magnitudes. They determined that $I$ is more or less identical to F814W, while $V$ is dependent on both F606W and the F606W--F814W color.

After applying the magnitude transformations we counted model field stars that fell within the $2\sigma_{c}$ envelope. We then scaled this number by WFC3's field of view to determine how many stars would be expected in our field. Hess diagrams of the Besan\c{c}on data along with the main sequence envelopes are shown in the left and center panels of Figure~\ref{fig:Hess}. The right panel shows the scaled field star contamination per magnitude bin. It is clear that the contamination is uniform across the main sequence and unlikely to exceed two stars per magnitude bin. 

\subsection{Blue Straggler Stars}\label{BSS}

Potential blue straggler stars (BSS) in each cluster were determined by eye. Given the scatter in the main sequence, the stars highlighted in blue in Figure~\ref{fig:CMD} represent a rudimentary estimate of the BSS populations. Fortunately, this is more than sufficient for our entirely qualitative purposes.

In Pal 13 we recover the substantial population of BSS noted by \citet{Borissova1997} and \citet{Siegel2001}, and studied in detail by \citet{Clark2004}. The high specific frequency of blue stragglers ($f_{\rm BS}$) coupled with Pal 13's low mass is consistent with the observed sub-linear relationship between the number of blue stragglers and cluster mass, which leads to a high $f_{\rm BS}$ in low-mass clusters \citep[e.g.][]{KLS2009,LSK2011,Leigh2013}. 

The CMD of AM 4 does not show this same high $f_{\rm BS}$. 

\subsection{Binary Stars}\label{bin}

The secondary sequences above the main sequences in Figure~\ref{fig:CMD} strongly suggests that each cluster contains a substantial number of unresolved binary stars. The binary stars of Pal 13 have been a topic of previous study, discussed by \citet{Clark2004}, \citet{Blecha2004}, \citet{Kupper2011} and \citet{Bradford2011}. The binary sequence in AM 4 is less pronounced, and has not been observable in previous CMDs. While a detailed analysis of the binary fraction (following the recent work of \citet{Milone2012b}) would be informative, it is beyond the scope of this paper. It is worth noting, however, that binary stars play a pivotal role in GC dynamical evolution. Therefore a substantial population of binary stars points to an interesting dynamical history.

\section{Results}\label{results}
\subsection{Physical Properties}\label{properties}
\begin{figure*}[t]
\begin{center}
\includegraphics[width = 6.5in]{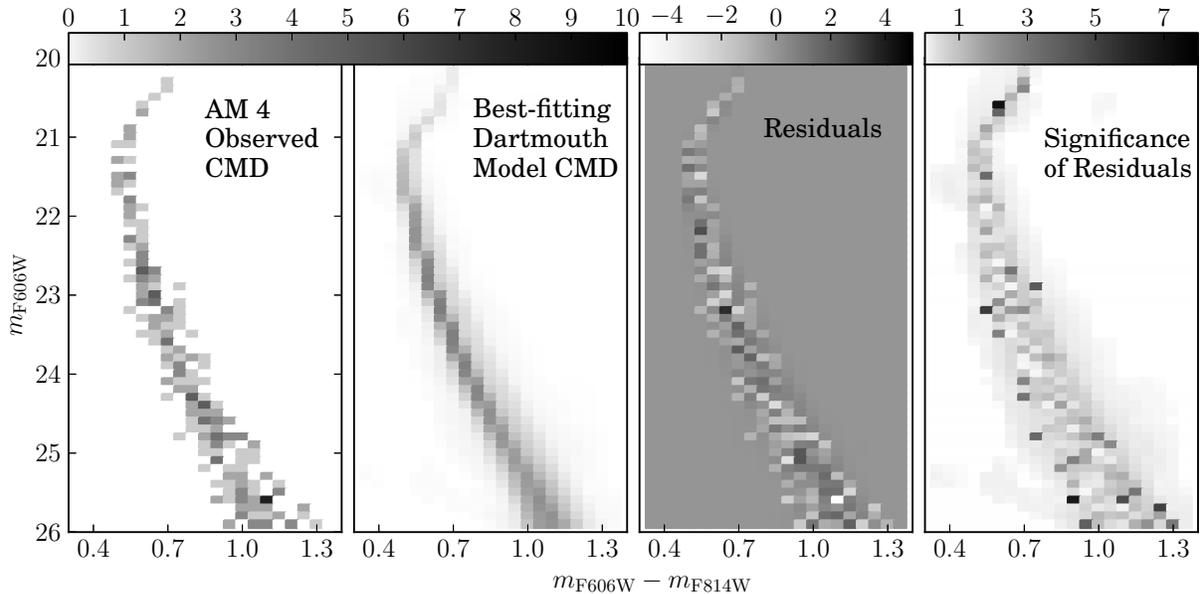}
\caption{Hess diagrams of the observed and best-fitting Dartmouth model CMDs of AM 4. The leftmost panel shows the observed CMD, disregarding blue stragglers and field stars while the panel second to the left shows the modeled CMD alone. The panel second to the right shows the difference between the observed and model CMDs in star counts, and the rightmost panel shows the difference between the observed and model CMDs scaled to the Poisson variance of the model CMD.}
\label{fig:am4_dart}
\end{center}
\end{figure*}

\begin{figure*}[t]
\begin{center}
\includegraphics[width = 6.5in]{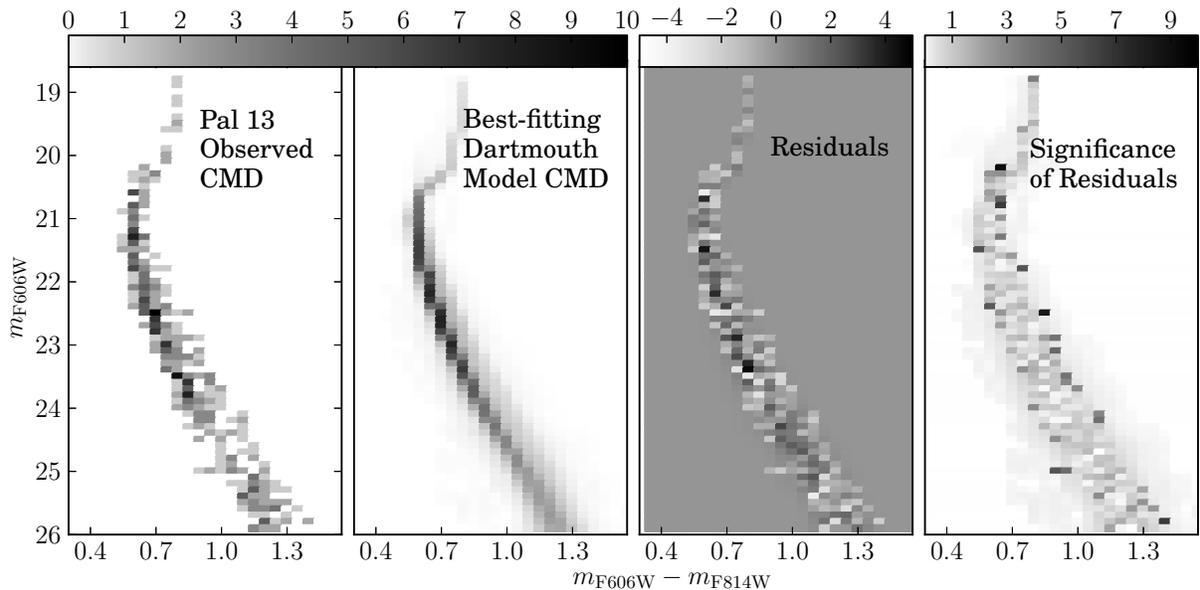}
\caption{Hess diagrams of the observed and best-fitting Dartmouth model CMDs of Pal 13. Panels are the same as in Figure~\ref{fig:am4_dart}.}
\label{fig:pal13_dart}
\end{center}
\end{figure*}

We determined properties of the clusters using the CMD fitting program MATCH \citep{dol02}. MATCH constructs synthetic CMDs of simple stellar populations from user-defined parameters including a stellar mass function (MF), binary fraction, a searchable range of distance and extinction values, fixed values of age, metallicity, and bins in color and magnitude. It then convolves the model CMD with observational biases as measured from the artificial star tests. MATCH computes the likelihood of the data given the model CMD using a Poisson likelihood statistic, enabling the characterization of the physical properties of a resolved stellar population. Although MATCH has primarily been used for analysis of field star populations \citep[e.g., dwarf galaxies in the Local Group and Local Volume,][]{dol05, tol09, wei11}, the underlying technique is readily adaptable to analysis of any resolved stellar population \citep[e.g.,][]{ski02, gal05}. 

\begin{deluxetable*}{ccr@{.}lr@{.}lr@{.}lc}
\tablewidth{0pt}
\tablecaption{Properties of AM 4 and Pal 13}
\tablehead{ Cluster &Property & \multicolumn{2}{c}{Dartmouth models} & \multicolumn{2}{c}{Padova models} & \multicolumn{2}{c}{BaSTI models} & \multicolumn{1}{c}{Literature values}} \\

\startdata
\multirow{5}{*}{AM 4} &
Age (Gyr) & $>13$&$0$ & $9$&$8\phantom{0} \pm 0.7$ & $>12$&$5$ & $9.0\phantom{0} \pm 0.5$\tablenotemark{a} \\
& $\log{Z/Z_\odot}$ & $-1$&$68 \pm 0.08$ & $-1$&$01 \pm 0.09$ & $-1$&$77 \pm 0.12$ & $\sim-0.97$\tablenotemark{a}  \\
& $(m-M)_V$ & $17$&$47 \pm 0.03$ & $17$&$53 \pm 0.04$ & $17$&$43 \pm 0.03$ & $17.7\phantom{0} \pm 0.2$\tablenotemark{a} \\
& $A_V$ & $0$&$19 \pm 0.02$ & $0$&$22 \pm 0.02$ & $0$&$19 \pm 0.02$ & $0.12 \pm 0.01$\tablenotemark{a} \\
& $\alpha$ & $+0$&$68 \pm 0.34$ & $+0$&$63 \pm 0.29$ & $+0$&$36 \pm 0.49$ & --- \\ 
\\
\\
\multirow{5}{*}{Pal 13} &
Age (Gyr) & $13$&$4 \pm 0.5$ & $>12$&$9$ & $>12$&$9$ & $12$\tablenotemark{b}\\
& $\log{Z/Z_\odot}$ & $-1$&$55 \pm 0.06$ & $-0$&$32 \pm 0.03$ & $-1$&$44 \pm 0.10$ & $-1.5 \pm 0.1\phantom{0}$\tablenotemark{b}  \\
& $(m-M)_V$ & $17$&$17 \pm 0.02$ & $16$&$83 \pm 0.02$ & $17$&$02 \pm 0.02$ & $16.93 \pm 0.10$\tablenotemark{c} \\
& $A_V$ & $0$&$43 \pm 0.01$ & $0$&$09 \pm 0.01$ & $0$&$34 \pm 0.01$ & $0.34 \pm 02$\tablenotemark{d} \\
& $\alpha$ & $-1$&$67 \pm 0.25$ & $-1$&$34 \pm 0.22$ & $-1$&$96 \pm 0.45$ & --- \\ 
\\
\hline
\multicolumn{9}{p{15cm}}{MF slope $\alpha$ given in the form $dN/dm \sim m^{-(1+\alpha)}$ (where $\alpha = +1.35$ for a Salpeter MF)}\\
\multicolumn{9}{p{15cm}}{a - \cite{Carraro2009}, b - \cite{Bradford2011}, c - \citet{Cote2002}, d - \citet{Schlegel1998}}
\enddata
\label{tab:physical}
\end{deluxetable*}

In this paper, we characterize the cluster properties using a power-law present day mass function (MF) with a mass range of 0.1 to 120 $M_\odot$ and a binary fraction of 0.35, where the mass of the secondary is drawn from a uniform mass distribution ranging from zero to the mass of the primary, and utilize Dartmouth, BaSTI, and Padova stellar evolution models \citep{Dotter2008a,mar08,gir10,pie04}. The use of multiple stellar models is particularly important for CMD analysis, as the systematic differences between stellar evolution libraries are frequently the dominant source of uncertainty in stellar population analysis \citep[e.g.,][]{wei11, dol12}.

We analyzed the clusters as follows. We first conducted a coarse grid search with resolution of 0.05 in $(m-M)_0$, $A_V$, and $\log(t)$; and 0.1 in log $Z/Z_\odot$ and MF slope; centered around the distance and extinction values listed in the \citet{Harris2010} catalog [AM4: $(m-M)_V = 17.69$, $A_V = 0.155$; Palomar 13: $(m-M)_V = 17.23$, A$_V = 0.155$]. These initial step sizes were chosen arbitrarily. The search box was iteratively modified until it was centered on the best-fitting solution. We only considered models of simple stellar populations, to the limits of MATCH's maximum resolution (0.05 dex in age, 0.1 dex in metallicity). Hess diagrams of observed and synthetic data were created with resolution 0.1 in magnitude and 0.05 in color.

For each cluster we computed likelihood values over the full coarse grid. The initial solutions indicated that the cluster parameters were constrained with degrees of precision better than the grid size, motivating us to re-run a finer optimized grid for each cluster. An optimized grid allows us to adequately sample the parameter space near the maximum likelihood. For this grid, we found a resolution of 0.02 mags in distance and extinction, 0.0167 dex in $\log(t)$ and 0.0333 dex in metallicity were fine enough to adequately sample the likelihood space. For MF slope we found it appropriate to increase the resolution to 0.2.

The process for converting the likelihoods of a grid of samples into characterizations of the parameters was done as follows. First, because MATCH's age and metallicity resolution prevented adequate sampling in those parameters, we created a supersampled grid of probability densities using a cubic spline to interpolate fit parameters at intermediate ages and metallicities. The probability density functions (PDFs) were then marginalized over each of the five axes and converted into cumulative probability distributions, with the 50th percentile point being reported as our best-fitting value while half the difference between the 16th and 84th percentile points is quoted as the uncertainty. The mean and sigma of the normal are reported for the MF. We note that the resulting measurements are consistent with the best-fitting models in all cases, but that the approach described here allows a slightly higher degree of precision in the identification of the 16th, 50th, and 84th percentile points.

\begin{figure}[t!]
\begin{center}
\includegraphics[width = 3in]{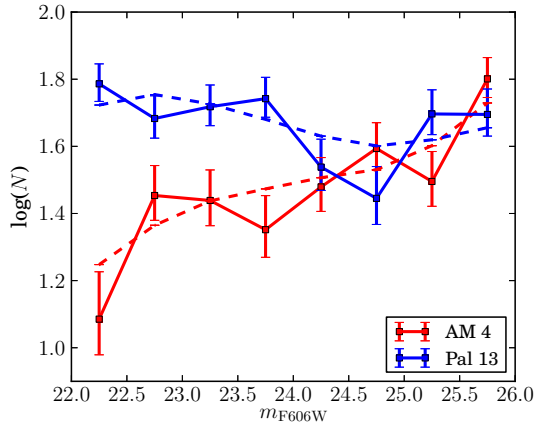}
\caption{The MSLFs of AM 4 (red) and Pal 13 (blue). Data (solid lines) is overlaid with the best-fitting model MSLF (dashed lines). The MSLFs have been corrected for completeness and field star contamination. It is clear that AM 4 has a significantly more bottom-heavy MSLF than Pal 13.}
\label{fig:LFs}
\end{center}
\end{figure}

As shown in Table \ref{tab:physical}, differences in parameters derived from different stellar models are much larger than the random uncertainties. This finding highlights the importance of using multiple stellar models when analyzing a stellar population. That is, for a given stellar model, the CMD contains sufficient information to provide precise constraints on each parameter. However, the differences in parameters derived with different stellar models indicates that the intrinsic accuracy of the stellar models is the dominant source of uncertainty. In our specific case, it is known that the Padova models produce a warmer/bluer red giant branch than other stellar models \citep[e.g.,][]{gal05, con10, wei12}. As a result, to match observed CMDs they typically require higher metallicities, which in turn affects other features such as the color and luminosity of the sub-giant branch. Such issues may be alleviated with the updated PARSEC models from the Padova group \citep{bre12}; however, they are not currently available for use in MATCH. In comparison, the Dartmouth models have been extensively calibrated using HST observations of Galactic globular clusters, and therefore may be better suited to globular cluster analysis. However, there are known shortcomings in the Dartmouth models, e.g., they produce blue horizontal branches for ages much younger than conventionally expected, leading to a different set of data-model mismatches \citep[e.g.,][]{dol12}. As the present data do not include horizontal branch stars, this issue does not concern us. More significant is that MATCH indicates that better solutions were obtained with the Dartmouth models than with the Padova and BaSTI models. We present Hess diagrams of the best-fitting Dartmouth solutions in Figures~\ref{fig:am4_dart} and \ref{fig:pal13_dart}, and the best-fitting model MSLF in Figure~\ref{fig:LFs}. A specific analysis of how the selected stellar models influence the characterization of globular clusters is beyond the scope of the present paper, and instead we simply re-emphasize the importance of including multiple stellar models in the analysis of any stellar population.

\subsection{Consistency Checks}\label{consistent}

As both Pal 13 and AM 4 have been observed before, it is instructive to compare the results of our CMD fitting to the properties derived by previous authors. As the Dartmouth models provide a better fit to the observed CMDs than either the Padova or BaSTI models, we use only properties derived using the Dartmouth models for comparison. The final column of Table~\ref{tab:physical} shows the age, metallicity, distance modulus and extinction as calculated by \citet{Carraro2009}, \citet{Bradford2011}, \citet{Cote2002} and \citet{Schlegel1998}. 

For AM 4, our age and metallicity measurements from MATCH differ from those published by \citet{Carraro2009}. The \citet{Carraro2009} ground-based data only extend $\sim2$ mag below the main-sequence turnoff and have large photometric errors. This, coupled with the fact that the overall mass of AM 4 is so low that it has no discernible red giant branch, makes the \citet{Carraro2009} age and metallicity estimates very uncertain. The depth of our HST photometry of AM 4 allows us to match model CMDs to the main sequence turnoff and to 4 magnitudes of the main sequence below the turnoff, and we are thus able to break the age-metallicity degeneracy that can plague ground-based studies of clusters without red giants. For Pal 13, the parameters we derive with MATCH agree well with those of \citet{Bradford2011}, \citet{Cote2002} and \citet{Schlegel1998}. It is particularly reassuring that our photometric metallicity estimate agrees with the spectroscopic metallicity from \citet{Bradford2011}.

As MF slopes have never been published for AM 4 or Pal 13, we computed the MF slopes implied by the literature and compared them to our results. To do this we used the Dartmouth model isochrone grid to generate isochrones with the previously published ages, metallicities and extinctions listed in Table~\ref{tab:physical}. We then converted these isochrones to MSLFs assuming a single-sloped power law MF of the form $dN/dm \sim m^{-(1+\alpha)}$ (where $\alpha = +1.35$ for a Salpeter MF). Our allowed slopes covered the range $-4 < \alpha < +4$. The resulting model MSLFs were scaled to match the number of stars in our cluster MSLFs, and compared using a standard $\chi^2$ statistic. This method gives a mean MF slope $\alpha = +0.72^{+0.46}_{-0.43}$ for AM 4 and $\alpha = -1.73^{+0.37}_{-0.39}$ for Pal 13. These are consistent with the MF slopes found by fitting the observed CMDs. 

Our final check was to compare the uncertainties on our MF slope measurements to the theoretical minimum ($\Delta\alpha$) proposed by \citet{wei13}. The mass ranges for the best-fitting Dartmouth solutions ($0.42M_\odot < M < 0.76M_\odot$ for AM 4 and $0.39M_\odot < M < 0.79M_\odot$ for Pal 13) are out of the regime in which $\Delta\alpha$ can be well modeled analytically. A qualitative estimate of the theoretical limit gives $\Delta\alpha \sim 0.4$, which is comparable to the uncertainties given by MATCH.

While not strictly a consistency check, we took a moment to consider the impact that unresolved binary stars may have had on our results. When running MATCH we assumed a binary fraction of 0.35 and a flat distribution of mass ratios ($0 < q < 1$). Both assumptions are in line with published binary fractions for Pal 13 (see references in Section~\ref{bin}) and other GCs of this size and magnitude \citep{Milone2012b}. It is possible that AM 4 has an unusual population of binaries, but it is unlikely that this has impacted our findings. Determination of the best-fitting model with MATCH is not a strong function of binary fraction, as any unresolved binary whose secondary noticeably affects the system color and magnitude will not lie on the single-mass main sequence and therefore is not a source of confusion.

\section{Discussion}\label{disc}

AM 4 and Pal 13 belong to a unique class of faint halo objects whose nature is not entirely understood. While mass loss has been both theorized and demonstrated in massive clusters (see references in the introduction to this paper), the role mass loss plays in these low-mass GCs is unclear. 

\begin{figure}[b]
\begin{center}
\includegraphics[width = 3.5in]{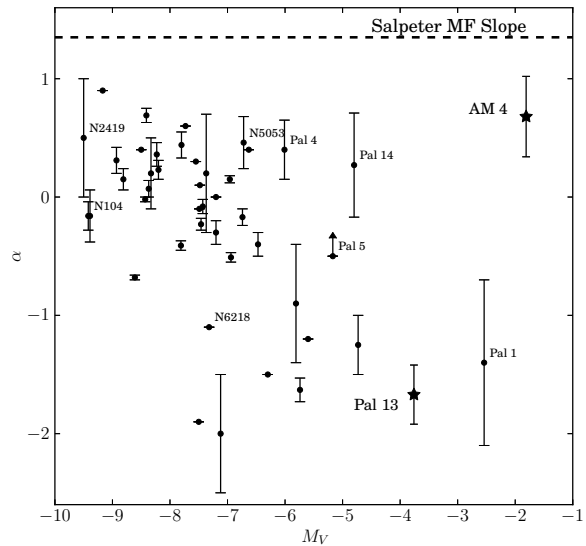}
\caption{Absolute visual magnitude versus MF slope (with the convention $dN/dm \sim m^{-(1+\alpha)}$) for AM 4, Pal 13, and 44 GCs from the literature. We have highlighted several of the clusters mentioned specifically in this paper. For AM 4 and Pal 13 we plot the MF slope given using the Dartmouth model isochrones, as they provided the best fit to the observed CMD. AM 4 is clearly anomalous both with respect to Pal 13 and with respect to other clusters of comparable luminosity.}
\label{fig:alpha}
\end{center}
\end{figure}

To put the MF slopes of AM 4 and Pal 13 into a broader context, we plot $\alpha$ versus $M_V$ for 44 additional GCs in Figure~\ref{fig:alpha}. The data for these GCs are presented in Table \ref{tab:alpha_lit}, and we have included error bars where provided in the literature. For each cluster, the MF covers the mass range $M_{min} < M \lesssim 0.8 M_\odot$, where $M_{\rm min}$ is listed in Table~\ref{tab:alpha_lit}.

\begin{deluxetable}{llr@{.}lcc}
\tablewidth{0pt}
\tablecaption{Mass Function Slope and Integrated Magnitudes} 
\tablehead{
\multicolumn{1}{c}{Object} & \multicolumn{1}{c}{$M_{\rm V}$} & \multicolumn{2}{c}{$\alpha$} & $M_{\rm min} (M_\odot)$ & Ref.}\\
\\

\startdata
AM 1 & $-4.73$ & $-1$&$25 \pm 0.25$ & -- & 6 \\ 
AM 4 & $-1.81$ & $+0$&$68 \pm 0.34$ & 0.39 & 16\\
NGC 104 & $-9.42$ & $-0$&$16 \pm 0.12$ & 0.2& 1 \\ 
NGC 288 & $-6.74$ & $-0$&$17 \pm 0.07$ & 0.2&1 \\ 
NGC 362 & $-8.41$ &  $+0$&$69 \pm 0.06$ & 0.2&1 \\ 
NGC 1261 & $-7.81$ & $-0$&$41 \pm 0.04$ & 0.2&1 \\ 
NGC 1851 & $-8.33$ & $+0$&$2 \pm 0.3$ & 0.5 & 14 \\
NGC 2298 & $-6.3$ & $-1$&$5$ & 0.3 & 10 \\ 
NGC 2419 & $-9.5$ & $+0$&$5 \pm 0.5$ & 0.5 & 3 \\
NGC 2808 & $-9.39$ & $-0$&$16 \pm 0.22$ & 0.3 & 15 \\ 
NGC 3201 & $-7.46$ & $-0$&$23 \pm 0.05$ & 0.2&1 \\ 
NGC 4590 & $-7.37$ & $+0$&$2 \pm 0.5$ & 0.5 & 11\\
NGC 5053 & $-6.72$ & $+0$&$46 \pm 0.22$ & 0.2&1 \\ 
NGC 5139 & $-10.29$ & $+0$&$2$ & 0.3 &10 \\ 
NGC 5272 & $-8.93$ & $+0$&$31 \pm 0.11$ & 0.2&1 \\ 
NGC 5286 & $-8.61$ & $-0$&$68 \pm 0.02$ & 0.2&1 \\ 
NGC 5466 & $-6.96$ & $+0$&$15 \pm 0.03$ & 0.2&1 \\ 
NGC 5904 & $-8.81$ & $+0$&$15 \pm 0.09$ & 0.2&1 \\ 
NGC 5927 & $-7.8$ & $+0$&$44 \pm 0.11$ & 0.2&1 \\ 
NGC 6093 & $-8.23$ & $+0$&$36 \pm 0.1$ & 0.2&1 \\ 
NGC 6121 & $-7.2$ & $+0$&$0$ & 0.3 &10 \\ 
NGC 6171 & $-7.12$ & $-2$&$0 \pm 0.5$ & 0.5 & 11\\
NGC 6205 & $-8.43$ & $-0$&$02 \pm 0.02$ & 0.2&1 \\ 
NGC 6218 & $-7.32$ & $-1$&$1$ & 0.3 &10 \\ 
NGC 6254 & $-7.48$ & $+0$&$1$ & 0.3 &10 \\ 
NGC 6341 & $-8.2$ & $+0$&$23 \pm 0.08$ & 0.2&1 \\ 
NGC 6352 & $-6.47$ & $-0$&$4 \pm 0.1$ & 0.3 & 5 \\ 
NGC 6362 & $-6.94$ & $-0$&$51 \pm 0.04$ & 0.2&1 \\ 
NGC 6366 & $-5.74$ & $-1$&$63 \pm 0.1$ & 0.2 & 13 \\
NGC 6397 & $-6.63$ & $+0$&$4$ & 0.3 &10 \\ 
NGC 6496 & $-7.2$ & $-0$&$3 \pm 0.1$ & 0.3 & 5 \\ 
NGC 6541 & $-8.37$ & $+0$&$07 \pm 0.07$ & 0.2&1 \\ 
NGC 6624 & $-7.49$ & $-0$&$1$ & 0.2 & 12 \\
NGC 6656 & $-8.5$ & $+0$&$4$ & 0.3 &10 \\ 
NGC 6712 & $-7.5$ & $-1$&$9$ & 0.3 &10 \\ 
NGC 6752 & $-7.73$ & $+0$&$6$ & 0.3 &10 \\ 
NGC 6809 & $-7.55$ & $+0$&$3$ & 0.3 &10 \\ 
NGC 6838 & $-5.6$ & $-1$&$2$ & 0.3 &10 \\ 
NGC 7078 & $-9.17$ & $+0$&$9$ & 0.3 &10 \\ 
NGC 7099 & $-7.43$ & $-0$&$08 \pm 0.06$ & 0.2&1 \\ 
NGC 7492 & $-5.81$ & $-0$&$9 \pm 0.5$ & 0.6 & 7 \\ 
Pal 1 & $-2.54$ & $-1$&$4 \pm 0.7$ & 0.65 &2 \\ 
Pal 4 & $-6.01$ & $+0$&$4 \pm 0.25$ & 0.55 &4 \\ 
Pal 5 & $-5.17$ & $\leq -0$&$5$ & 0.3 & 8 \\
Pal 13 & $-3.76$ & $-1$&$67 \pm 0.25$ & 0.42 & 16 \\ 
Pal 14 & $-4.8$ & $+0$&$27 \pm 0.44$ & 0.53 & 9 \\ 
\hline
\\
\multicolumn{6}{p{7cm}}{References -- (1) \citet{Paust2010}, (2) \citet{Rosenberg1998}, (3) \citet{Bellazzini2012}, (4) \citet{Frank2012}, (5) \citet{Pulone2003}, (6) \citet{Dotter2008b}, (7) \citet{Cote1991}, (8) \citet{GrillmairSmith2001}, (9) \citet{Jordi2009}, (10) \citet{deMarchi2007}, (11) \citet{Capaccioli1991} (12) \citet{Grabhorn1991} (13) \citet{Paust2009} (14) \citet{Saviane1998} (15) \citet{Milone2012a} (16) This work}
\enddata
\label{tab:alpha_lit}
\end{deluxetable}

Calculating the Pearson correlation coefficient $r^2$ reveals a mild correlation between $\alpha$ and $M_V$: $r^2 = 0.28 \pm 0.07$ including Pal 13, and $r^2 = 0.14 \pm 0.05$ with AM 4 and Pal 13. We estimated the uncertainties on these correlations due to the uncertainties on the MF slopes using a Monte Carlo simulation with 50,000 realizations. This is a substantially more significant correlation than \citet{Paust2010} found ($r^2 = 0.01 \pm 0.05$), as their data only include GCs with magnitudes in the range $-9.42 < M_{\rm V} < -6.72$.  

This correlation may indicate that low mass GCs have lost a considerable amount of mass. However, Pal 4, Pal 14 and AM 4 do not follow this trend. AM 4 in particular is much steeper than both Pal 13 and other comparably faint clusters. Refocusing on Pal 13 and AM 4, we have investigated a variety of physical properties, none of which appear to have a large impact on the clusters' mass-loss histories. In particular, our data do not confirm the correlation between Galactic location and MF slope found by \citet{Djorgovski1993} and \citet{Piotto1999}, as our distance moduli give a Galactocentric radius $R_{\rm GC} 
\sim 24.0$ kpc for AM 4 and $R_{\rm GC} \sim 23.5$ kpc for Pal 13 (assuming $R_\odot = 8.5$ kpc). 

We have not, so far, examined the orbital dynamics of these clusters. While the current position in the Galaxy appears to have little impact on the MF, the properties of a GC's orbit (inclination, eccentricity, period, etc.) determine the rate at which the cluster loses stars to processes such as two-body relaxation, tidal stripping and gravitational shocking, \citep[e.g.][]{Fall2001,Lamers2010}. As the rate of mass loss through gravitational stocking and tidal stripping increases in regions of higher density or stronger tidal field, eccentricity and perigalactic distance are two orbital parameters that are particularly relevant.

While its orbital phase is still debated, Pal 13 has been established as having an inclined, highly eccentric orbit. \citet{Siegel2001} find eccentricity $e = 0.76$ and perigalacticon at $R_p = 11.2$ kpc, while \citet{Kupper2011} find $e = 0.83$ and $R_p = 3.5$ kpc. Hence, it is likely that Pal 13 has been subjected to the sorts of processes that strip low mass stars. It is possible that AM 4 has simply not been comparably stripped, and thus has a MSLF/MF with more faint stars. Without similar velocity data for AM 4, however, we cannot confirm this explanation.

A competing effect is that not all populations necessarily have the same initial (i.e. primordial) mass function (IMF). \citet{Carraro2009} suggested that AM 4 may be associated with the Sagittarious dwarf spheroidal (Sgr dSph) galaxy rather than the Milky Way. While our work can not confirm this suggestion, it is not unreasonable to suggest that IMFs of Sgr GCs differ from those of MW clusters. However, if we look at other Sgr GCs \citep{Law2010} we see that NGC 5053 (labeled on Figure~\ref{fig:alpha}) does not have a steeper MF than similar MW GCs \citep{Paust2010}, and the LF of Whiting 1 was found to be remarkably flat \citep{Carraro2007}. In addition, the MF of the Sgr dSph itself is flatter than the MW \citep{Geha2013}, and there is no evidence to suggest that dSphs with flatter MFs would have GCs with steeper MFs than their MW counterparts. We conclude that AM 4's possible affiliation with the Sgr dSph is unlikely to have any bearing on its IMF.

Putting AM 4's affiliations aside, there has been evidence for multiple IMFs in local group clusters \citep{Zaritsky2012,Zaritsky2013}. However, these authors have found that IMFs differ between old, metal-poor clusters and young, metal-rich clusters. While an age of $\sim10$ Gyr does make AM 4 several Gyr older than the other very faint GCs, it is the same age as Pal 13. In addition, while our analysis indicates that the metallicities of AM 4 and Pal 13 are different, neither constitutes a metal-rich cluster. It is thus unlikely that they are disparate enough in age and metallicity to have two different IMFs on the basis of \citet{Zaritsky2012,Zaritsky2013}.

In conclusion, we find that Pal 13 displays all of the characteristics of a GC that has lost a considerable amount of mass. Fitting Pal 13 into the large picture of GCs suggests that MF slope is correlated with cluster magnitude. However, AM 4 complicates this picture as it does not appear to have the same mass-loss history as the other extremely faint GCs (e.g. Pal 13, Pal 1). It is possible that, along with Pal 4 and Pal 14, AM 4 belongs to a subset of clusters that have escaped the tidal processes that have affected other clusters. Velocity and proper motion measurements of AM 4, as well as measurements of the MF slopes of other faint GCs (i.e., Koposov 1 and 2, E3) will help answer these questions in the future.

\acknowledgements

We would like to thank Jason Kalirai for his invaluable assistance with photometry and data reduction, and Alis Deason, Claire Dorman, Brad Holden and Connie Rockosi for useful discussions. We would also like to thank Namita Ravi for her work tracking down mass function slopes. Finally, we thank the anonymous referee for his or her helpful comments.

This work was supported by the award STScI GO-11680.01-A. K.H. and P.G. acknowledge support from NSF grant AST-1010039 and NASA grant HST-GO-12055. 
 
\bibliographystyle{apj}
\bibliography{references}

\end{document}